\documentclass[aps,showpacs,prd,twocolumn]{revtex4}%
\usepackage{graphicx}
\usepackage{amssymb}
\usepackage{amsmath}
\usepackage{color}
\usepackage{float}
\usepackage{ulem}
\usepackage{accents}
\usepackage{graphicx}
\usepackage{graphicx}
\usepackage{amsfonts}
\usepackage{slashed}
\usepackage[colorlinks=true, pdfstartview=FitV,
linkcolor=blue,citecolor=blue,urlcolor=blue, breaklinks=true]{hyperref}%
\providecommand{\U}[1]{\protect\rule{.1in}{.1in}}
\begin{document}
\title{\textbf{Constraining CPT-odd nonminimal interactions in the Electroweak
sector}}
\author{V. E. Mouchrek-Santos}
\email{victor\_mouchrek@hotmail.com }
\author{Manoel M. Ferreira Jr}
\email{manojr.ufma@gmail.com}
\affiliation{Departamento de F\'{\i}sica, Universidade Federal do Maranh\~{a}o, Campus
Universit\'{a}rio do Bacanga, S\~{a}o Lu\'{\i}s - MA, 65080-805 - Brazil}

\begin{abstract}
In this work, we propose two possibilities of CPT-odd and Lorentz-violating
(LV) nonminimal couplings in the Electroweak sector. These terms are
gauge-invariant and couple a fixed 4-vector to the physical fields of the
theory. After determining the LV contributions to the electroweak currents, we
reassess the evaluation of the decay rate for the vector mediators $W$ and
$Z$. Using the experimental uncertainty in these decay rates, upper bounds of
$1$ part in $10^{-15}$ (eV )$^{-1}$ and $10^{-14}$ (eV )$^{-1}$ are imposed on
the magnitude of the proposed nonmimal interactions.

\end{abstract}

\maketitle

\section{Introduction}

Mechanisms of spontaneous Lorentz violation have been proposed in some
candidate theories of quantum gravity. As a consequence, Lorentz-violating
(LV) background\textbf{ }tensors (generated as vacuum expectation values) are
coupled to the physical fields of the Standard Model. The most general
effective theory considering the explicit breaking of Lorentz and CPT symmetry
is the minimal Standard Model Extension (mSME) \cite{Colladay}, which is an
extension of the Standard Model, $SU\left(  3\right)  \times SU\left(
2\right)  \times U\left(  1\right)  $, featuring terms breaking Lorentz and
CPT symmetries in all of its sectors: lepton, quark, Yukawa, Higgs and gauge.
Investigation of Lorentz symmetry violation is a rich line of research,
embracing developments in the electromagnetic sector \cite{KM1},
\cite{photons1}, fermion sector \cite{fermion}, including photon-fermion
interactions \cite{Vertex},\cite{Vertex2},\cite{Carone}. Such studies have
detailed LV effects in very distinct physical systems, allowing to construct a
precision programme to determine to what extent the Lorentz covariance is
preserved in nature (by means of tight upper bounds on the LV coefficients).
Nonminimal LV interactions have also been examined in an extension of the
SME\ encompassing higher-order derivatives in both the gauge \cite{NMSME} and
the fermion sector \cite{NMSME2}. Some other models containing
higher-dimension operators \cite{Reyes} and higher derivatives \cite{HD} have
also been proposed and developed.

In the electroweak sector of the mSME \cite{Colladay}, the SU(2) and U(1)
gauge fields are properly coupled to LV fixed tensors in renormalizable
dimension four terms. The mSME lepton sector is composed of a CPT-even and a
CPT-odd term, that is,
\begin{align}
\mathcal{L}_{lepton}^{even}  &  =i(c_{L})_{\mu\nu AB}\bar{L}_{A}\gamma^{\mu
}iD^{\nu}L_{B}+i(c_{R})_{\mu\nu AB}\bar{R}_{A}\gamma^{\mu}iD^{\nu}%
R_{B},\label{Leptoneven}\\
\mathcal{L}_{lepton}^{odd}  &  =-(a_{L})_{\mu AB}\bar{L}_{A}\gamma^{\mu}%
L_{B}-(a_{R})_{\mu AB}\bar{R}_{A}\gamma^{\mu}R_{B},
\end{align}
where $A,B=1,2,3$ are the lepton flavor labels. At the same way, the SU(2) and
U(1) gauge sectors are modified by the CPT-even terms:%
\begin{equation}
\mathcal{L}_{gauge}^{even}=-\frac{1}{2}(k_{W})^{\mu\nu\alpha\beta}(W_{\mu\nu
}^{a}W_{\alpha\beta}^{a})-\frac{1}{4}(k_{B})_{\mu\nu\alpha\beta}B^{\mu\nu
}B^{\alpha\beta}, \label{LVgauge}%
\end{equation}
while the CPT-odd generate instabilities in the theory, and are not
considered. These coefficients are real, dimensionless, and possess the same
symmetries of the Riemann tensor. The pure CPT-even Higgs sector is also
modified by the following term:%

\begin{align}
&  \left.  \mathcal{L}_{Higgs}^{even}=\frac{1}{2}(k_{\phi\phi})_{\mu\nu
}\left(  D^{\mu}\phi^{a}\right)  ^{\dag}\left(  D^{\nu}\phi^{a}\right)
+h.c.\right. \nonumber\\
&  \left.  -\frac{1}{2}(k_{\phi B})_{\mu\nu}\phi^{a\dag}\phi^{a}B^{\mu\nu
}-\frac{1}{2}(k_{\phi W})^{\mu\nu}(\phi^{\dag}\times\phi)^{a}W_{\mu\nu}%
^{a},\right.  \label{Higgseven}%
\end{align}
while the Higgs CPT-odd has the form $i(k_{\phi})_{\mu}\left(  \phi
^{a}\right)  ^{\dag}\left(  D^{\mu}\phi^{a}\right)  ,$ with $(k_{\phi})_{\mu}$
having dimension of mass.

In the electroweak sector, LV studies were initially developed in connection
with meson decays $\left(  \pi^{-}\rightarrow\mu^{-}+\bar{\nu}_{\mu}\right)
$,\textbf{ }where the LV effects were considered at the level of the Feynman
propagator of the $W$ boson \cite{Pion}, $\left\langle W^{\mu\dagger}W^{\nu
}\right\rangle =-i\left(  g^{\mu\nu}+\chi^{\mu\nu}\right)  /M_{W}^{2},$ with
contributions coming from the Higgs $\left(  \phi\right)  $ and the $W$
sectors: $\chi^{\mu\nu}=k_{\phi\phi}^{\mu\nu}-\frac{i}{2g}k_{\phi W}^{\mu\nu
}+k_{W}^{\alpha\mu\beta\nu}p_{\alpha}p_{\beta}.$ Comparison with experimental
data led to upper bounds of 1 part in $10^{4}.$ Contributions of the $k_{W}$
coefficients (\ref{LVgauge}) to the\textbf{ }$W$ propagator, $\left\langle
W^{\mu\dagger}W^{\nu}\right\rangle =-i\left(  g^{\mu\nu}+\chi^{\mu\nu}\right)
/M_{W}^{2},$ jointly with contributions stemming from the Higgs sector,
$k_{\phi\phi},k_{\phi W},$ see Eq. (\ref{Higgseven}), were more explicitly
considered in Ref. \cite{LVBeta}, with implications on the allowed nuclear
decays and forbidden $\beta$ decays. This framework was also used: (i) to
reinterpret experiments dedicated to searching for preferred directions in
forbidden $\beta-$decays, implying upper bounds as tight as $10^{-8}$ on the
LV\ parameters \cite{LVBeta2}; (ii) to constrain $\beta$ decay rate
asymmetries to the level of 1 part in $10^{6}$ \cite{LVBeta3}; (iii) to study
isotopes that undergo orbital electron capture \cite{LVBeta4}; (iv) to analyse
LV effects on the kaon decay and evaluate asymmetries in the respective
lifetime \cite{Kaon}. Another interesting study considered LV coefficients
$(c_{L})_{\mu\nu AB}$ of the lepton sector (\ref{Leptoneven}), with the same
flavor ($A=B$),
\begin{align}
\mathcal{L}_{lepton}  &  =ic_{\alpha\beta}[i\bar{\psi}\gamma^{\alpha}%
\partial^{\beta}\psi+i\bar{\psi}_{\nu}\gamma^{\alpha}\partial^{\beta}\psi
_{\nu}\nonumber\\
&  +\bar{\psi}_{(L)}\gamma^{\alpha}W^{\beta(-)}\psi_{\nu(L)}+i\bar{\psi}%
_{\nu(L)}\gamma^{\alpha}W^{\beta(+)}\psi_{(L)}],
\end{align}
where $\psi,\psi_{\nu},\psi_{(L)}$ represent leptons, neutrinos and
left-handed leptons (of a given flavor), to examine effects on the pion-decay
rate \cite{Piond}, attaining upper bounds of the level of $10^{-4}.$ Some
works also examined the possibility of LV electroweak terms to make feasible
forbidden processes ($Z_{0}\longrightarrow\gamma+\gamma)$ \cite{EW2} or modify
the reactions such as $\gamma+e\rightarrow W+\nu_{e}$, $\gamma+\gamma
\rightarrow W+W$ \cite{EW2b}.\ Lepton flavor violating decays triggered by
renormalizable and nonrenormalizable (dimension five) terms belonging to the
Higgs sector were recently considered as well\cite{EW3}. Tree-level Z-boson
contributions to the polarized M\"{o}ller scattering were carried out,
allowing to improve $k_{W}$ upper bounds by two orders of magnitude
\cite{Hao}. Lorentz violation influence on neutrino oscillations was also
probed using a distinct framework \cite{Diaz}.

A dimension five LV\ nonminimal coupling (NMC), representing unusual
interactions between fermions and photons, $g^{\nu}\bar{\psi}\gamma^{\mu
}\gamma_{5}\tilde{F}_{\mu\nu}\psi$, was first introduced by means of the
derivative, $D_{\mu}=\partial_{\mu}+ieA_{\mu}+i\frac{\lambda}{2}\epsilon
_{\mu\lambda\alpha\beta}V^{\lambda}F^{\alpha\beta}$, in the Dirac equation
\cite{NM1}, where $V^{\mu}$ can be identified with the Carroll-Field-Jackiw
four-vector. Such a coupling has been addressed in numerous aspects
\cite{NMmaluf}, including the radiative generation of $CPT$-odd LV terms
\cite{Radio}, topological phases \cite{NMABC}, \cite{NMbakke}, and generation
of electric dipole moment \cite{Pospelov}. Dimension-five $CPT$-even NMCs were
also proposed in the context of the Dirac equation \cite{FredeNM1},
\cite{Jonas1}, \cite{Jonas2}, with MDM and EDM experimental measurements being
used to state upper bounds at the level of $1$ part in $10^{20}$ (eV)$^{-1}$
and $10^{24}$ (eV)$^{-1}$, respectively. A systematic investigation on NMCs of
dimension five and six was recently proposed in Ref. \cite{Koste2}.

Nonminimal interactions have been a topical issue in the latest years, mainly
in the fermion and electromagnetic sectors. However, a NMC in the lepton
electroweak sector of the Standard Model has not been proposed yet. In this
work, we introduce two possibilities of CPT-odd LV nonminimal interactions in
the Electroweak sector, the first one being proposed in the $U\left(
1\right)  _{Y}$ sector of the GSW model, while the second is considered in its
$SU\left(  2\right)  _{L}$ sector, both as extensions of the covariant
derivative. After determining the terms engendered in the interaction
Lagrangian, we evaluate the Lorentz-violating corrections to the decay rates
of the following mediators: $Z_{0}\rightarrow\bar{l}+l$ and $W^{-}\rightarrow
l+\bar{\nu}_{l}$. Using these results and the experimental uncertainty in the
measurements, we impose upper limits on the magnitude of the LV parameters,
achieving upper bounds as tight as $10^{-15}$ (eV)$^{-1}.$

\section{Basics about\textbf{ the GSW} model}

In the Glashow-Salam-Weinberg electroweak model (GSW), with a $SU\left(
2\right)  _{L}\times U\left(  1\right)  _{Y}$ gauge structure spontaneously
broken via the Higgs mechanism, the vector bosons, $W^{\pm}$, $Z^{0}$ and
$\gamma$ are mediator of the interactions, being introduced via minimal
coupling to the matter fields. In this theory, left-handed leptons ($L_{l})$
are represented by isodublets%
\begin{equation}
L_{l}=%
\begin{bmatrix}
\psi_{\nu_{l}}\\
\psi_{l}%
\end{bmatrix}
_{L}=\frac{1-\gamma_{5}}{2}%
\begin{bmatrix}
\psi_{\nu_{l}}\\
\psi_{l}%
\end{bmatrix}
, \label{LGSW1}%
\end{equation}
while right-handed leptons ($R_{l})$ are isosinglets,%
\begin{equation}
R_{l}=\left(  \psi_{l}\right)  _{R}=\left(  \frac{1+\gamma_{5}}{2}\right)
\psi_{l}, \label{LGSW2}%
\end{equation}
and $l=1,2,3$ is the lepton flavor label: $\psi_{l}=(e,\mu,\tau).$ The part of
the electroweak Lagrangian, in which the leptons interact directly with the
gauge fields, is $\mathcal{L}_{EW}=\mathcal{L}_{gauge}+\mathcal{L}_{lepton}%
,$\textbf{ }where\textbf{ }%

\begin{align}
\mathcal{L}_{gauge}  &  =-\frac{1}{4}\mathbf{W}_{\mu\nu}\cdot\mathbf{W}%
^{\mu\nu}-\frac{1}{4}B_{\mu\nu}B^{\mu\nu},\label{LGSW.14}\\
\mathcal{L}_{lepton}  &  =\bar{L}_{l}\gamma^{\mu}iD_{\mu}L_{l}+\bar{R}%
_{l}\gamma^{\mu}iD_{\mu}R_{l}, \label{LGSW.14b}%
\end{align}
with $\mathbf{W}_{\mu}=(W_{\mu}^{1},W_{\mu}^{2},W_{\mu}^{3})$ being a
four-vector gauge field which is a three-vector in isospin space, and $B_{\mu
}$ a gauge four-vector field, whose field strengths are
\begin{align}
\mathbf{W}_{\mu\nu}  &  =\partial_{\mu}\mathbf{W}_{\nu}-\partial_{\nu
}\mathbf{W}_{\mu}+g\mathbf{W}_{\mu}\times\mathbf{W}_{\nu},\label{LGSW10}\\
B_{\mu\nu}  &  =\partial_{\mu}B_{\nu}-\partial_{\nu}B_{\mu}. \label{Bmini}%
\end{align}
The covariant derivative involves both gauge fields,%

\begin{equation}
D_{\mu}=\partial_{\mu}-ig\mathbf{T\cdot W}_{\mu}-i\frac{g^{\prime}}{2}YB_{\mu
}. \label{CD1}%
\end{equation}
Here, $\mathbf{T}=\left(  T_{1},T_{2},T_{3}\right)  $ stands for the
generators of the group $SU\left(  2\right)  _{L}$, and $Y$ is the generator
of $U\left(  1\right)  _{Y}$ group, fulfilling $\left[  T_{i},T_{j}\right]
=i\varepsilon_{ijk}T_{k}$ and $\left[  T_{i},Y\right]  =0.$ Furthermore, $Y$
$=-1$ or $Y=-2$ for left-handed and right-handed leptons,
respectively\textbf{.} The lepton Lagrangian (\ref{LGSW.14b}) can be written
as $\mathcal{L}=i\bar{L}_{l}\gamma^{\mu}\partial_{\mu}L_{l}+i\bar{R}_{l}%
\gamma^{\mu}\partial_{\mu}R_{l}+\mathcal{L}_{int}^{\left(  l\right)  },$ with
the interaction part given as
\begin{align}
\mathcal{L}_{int}^{\left(  l\right)  }  &  =\frac{g}{2\sqrt{2}}\left(
J_{+}^{\left(  l\right)  \alpha}W_{\alpha}^{\left(  +\right)  }+J_{-}^{\left(
l\right)  \alpha}W_{\alpha}^{\left(  -\right)  }+J_{0}^{\left(  l\right)
\alpha}Z_{\alpha}\right) \nonumber\label{currents22}\\
&  -eJ_{EM}^{\left(  l\right)  \alpha}A_{\alpha},
\end{align}
where there appear charged currents, $J_{+}^{\left(  l\right)  \alpha}%
,J_{-}^{\left(  l\right)  \alpha}$, a neutral current, $J_{0}^{\left(
l\right)  \alpha},$ and the electromagnetic current, $J_{EM}^{\left(
l\right)  \alpha},$ given as
\begin{align}
J_{+}^{\left(  l\right)  \alpha}  &  =2\bar{L}_{l}\gamma^{\alpha}T_{+}%
L_{l}=\bar{\psi}_{\nu_{l}}\gamma^{\alpha}\left(  1-\gamma_{5}\right)  \psi
_{l},\\
J_{-}^{\left(  l\right)  \alpha}  &  =2\bar{L}_{l}\gamma^{\alpha}T_{-}%
L_{l}=\bar{\psi}_{l}\gamma^{\alpha}\left(  1-\gamma_{5}\right)  \psi_{\nu_{l}%
},
\end{align}%
\begin{align}
J_{0}^{\left(  l\right)  \alpha}  &  =\left(  \sqrt{2}\cos\theta\right)
^{-1}\left[  \bar{\psi}_{\nu_{l}}\gamma^{\alpha}\left(  1-\gamma_{5}\right)
\psi_{\nu_{l}}\right. \nonumber\\
&  \left.  -\bar{\psi}_{l}\gamma^{\alpha}\left(  g_{V}^{\prime}-g_{A}^{\prime
}\gamma_{5}\right)  \psi_{l}\right]  ,\label{currents.10}\\
J_{EM}^{\left(  l\right)  \alpha}  &  =-\left[  \bar{L}_{l}\gamma^{\alpha
}\left(  \frac{g^{\prime}\cos\theta}{2}-g\sin\theta T_{3}\right)  L_{l}\right.
\nonumber\\
&  \left.  +g^{\prime}\cos\theta\bar{R}_{l}\gamma^{\alpha}R_{l}\right]  .
\end{align}
Here, $\theta$ is the Weinberg angle, and $g,g^{\prime}$ are the coupling
constants, and the vector-axial interaction is controlled by $g_{A}^{\prime
}=1,$ \ $g_{V}^{\prime}=1-4\sin^{2}\theta.$ In the electroweak theory, the
photon ($A_{\mu})$ is a combination of the fields $W_{\mu}^{3}$ and $B_{\mu},$
that is, $A_{\mu}=B_{\mu}\cos\theta+W_{\mu}^{3}\sin\theta,$ while the neutral
intermediate boson is $Z_{\mu}=-B_{\mu}\sin\theta+W_{\mu}^{3}\cos\theta.$ The
inverse relations are also well known, $B_{\mu}=\cos\theta A_{\mu}-\sin\theta
Z_{\mu},W_{\mu}^{3}=\sin\theta A_{\mu}+\cos\theta Z_{\mu}.$ The generators and
isovector can be also written as $\mathbf{T=(}T_{+},T_{3},T_{-}),$
$\mathbf{W}_{\alpha}=(W_{\alpha}^{\left(  +\right)  }/\sqrt{2},W_{\mu}%
^{3},W_{\alpha}^{\left(  -\right)  }/\sqrt{2}),$ where $T_{\pm}=\sigma
_{x}/2\pm i\left(  \sigma_{y}/2\right)  ,$ $T_{3}=\sigma_{z}/2,$ and $W_{\mu
}^{\left(  \pm\right)  }=\frac{1}{\sqrt{2}}\left(  W_{\mu}^{1}\mp iW_{\mu}%
^{2}\right)  ,$ and $\sigma_{x},\sigma_{y},\sigma_{z}$ are the Pauli matrices.

\section{\ A nonminimal \textbf{coupling in the U}$\left(  1\right)  _{Y}$
sector of the GSW model}

We have already mentioned how Lorentz-violating terms are inserted in the mSME
electroweak sector. Another route to consider it involves higher dimensional,
nonrenormalizable NM operators. Gauge invariant NM interactions in the
electroweak sector can be proposed in the context of the covariant derivative
(\ref{CD1}). A first possibility, in the $U\left(  1\right)  _{Y}$ sector of
the GSW model,\textbf{ }is the NM derivative%
\begin{equation}
D_{\mu}=\partial_{\mu}-ig\mathbf{T\cdot W}_{\mu}-i\frac{g^{\prime}}{2}YB_{\mu
}+ig_{2}^{\prime}YB_{\mu\nu}C^{\nu}, \label{Bmn1}%
\end{equation}
where $C^{\nu}$ is a fixed 4-vector that establishes a preferred direction in
spacetime and violates Lorentz symmetry. Replacing such a derivative in
Lagrangian (\ref{LGSW.14b}), the nonminimal coupling yields additional
electromagnetic and neutral Lorentz-violating interactions,
\begin{equation}
\mathcal{L}_{LV\left(  1\right)  }=J_{EM\left(  LV\right)  }^{\left(
l\right)  \nu}A_{\nu}+J_{0\left(  LV\right)  }^{\left(  l\right)  \nu}Z_{\nu},
\end{equation}
given explicitly as%
\begin{gather}
J_{EM\left(  LV\right)  }^{\left(  l\right)  \nu}=\frac{g_{2}^{\prime}}{2}%
\cos\theta\left[  \bar{\psi}_{\nu_{l}}\gamma^{\mu}\left(  1-\gamma_{5}\right)
\psi_{\nu_{l}}\right]  C^{\nu}\partial_{\mu}\nonumber\\
-\frac{g_{2}^{\prime}}{2}\cos\theta\left[  \bar{\psi}_{\nu_{l}}\gamma^{\nu
}\left(  1-\gamma_{5}\right)  \psi_{\nu_{l}}\right]  C^{\mu}\partial_{\mu
}\nonumber\\
+g_{2}^{\prime}\cos\theta\lbrack j_{1}^{\mu}C^{\nu}\partial_{\mu}%
]-g_{2}^{\prime}\cos\theta\left[  j_{1}^{\nu}C^{\mu}\partial_{\mu}\right]  ,
\end{gather}%
\begin{gather}
J_{0\left(  LV\right)  }^{\left(  l\right)  \nu}=-\frac{g_{2}^{\prime}}{2}%
\sin\theta\left[  \bar{\psi}_{\nu_{l}}\gamma^{\mu}\left(  1-\gamma_{5}\right)
\psi_{\nu_{l}}\right]  C^{\nu}\partial_{\mu}\nonumber\\
+\frac{g_{2}^{\prime}}{2}\sin\theta\left[  \bar{\psi}_{\nu_{l}}\gamma^{\nu
}\left(  1-\gamma_{5}\right)  \psi_{\nu_{l}}\right]  C^{\mu}\partial_{\mu
}\nonumber\\
-g_{2}^{\prime}\sin\theta\left[  j_{1}^{\mu}C^{\nu}\partial_{\mu}\right]
+g_{2}^{\prime}\sin\theta\left[  j_{1}^{\nu}C^{\mu}\partial_{\mu}\right]  .
\label{LVJ0}%
\end{gather}
with $j_{1}^{\mu}(x)=\bar{\psi}_{l}\left(  x\right)  \gamma^{\mu}\left(
3+\gamma_{5}\right)  \psi_{l}\left(  x\right)  /2.$These expressions are
useful to show the processes that are directly affected, at tree-level, by the
nonminimal derivative\ (\ref{Bmn1}). We now examine the effect of this
nonminimal coupling on the decay of the $Z_{0}$ mediator in a pair lepton and
antilepton,
\begin{equation}
Z_{0}\rightarrow\bar{l}+l, \label{process1}%
\end{equation}
evaluating the contributions implied to the decay rate. The total neutral
current, $\left(  J_{0}^{\left(  l\right)  \mu}+J_{0\left(  LV\right)
}^{\left(  l\right)  \mu}\right)  Z_{\mu}$, that contributes for this process is%

\begin{gather}
=-\frac{g}{4\cos\theta}\bar{\psi}_{l}\left(  x\right)  \gamma^{\mu}\left(
g_{V}^{\prime}-\gamma_{5}\right)  \psi_{l}\left(  x\right)  Z_{\mu}\left(
x\right) \nonumber\\
-g_{2}^{\prime}\sin\theta\left[  j_{1}^{\eta}C^{\mu}\partial_{\eta}Z_{\mu
}\left(  x\right)  \right]  +g_{2}^{\prime}\sin\theta\left[  j_{1}^{\mu
}C^{\lambda}\partial_{\lambda}Z_{\mu}\left(  x\right)  \right]  , \label{TC1}%
\end{gather}
where the first term is the usual Lorentz invariant contribution, the second
and third terms stem from Eq.\ (\ref{LVJ0}). Expression (\ref{TC1})
shows\textbf{ }how the NMC \ (\ref{Bmn1}) affects\textbf{ }the vertex of the
neutral interaction. The scattering matrix for such a process is%
\begin{equation}
S=-i\int d^{4}x\left(  J_{0}^{\left(  l\right)  \mu}+J_{0\left(  LV\right)
}^{\left(  l\right)  \mu}\right)  Z_{\mu}=S_{0}+S_{LV\left(  1\right)
}+S_{LV\left(  2\right)  }, \label{VSL1.11}%
\end{equation}
where the zero order and first order contributions in the LV parameters are%
\begin{align}
S_{0}  &  =i\frac{g}{4\cos\theta}\int d^{4}x\text{ }\bar{\psi}_{l}\left(
x\right)  \gamma^{\mu}\left(  g_{V}^{\prime}-\gamma_{5}\right)  \psi
_{l}\left(  x\right)  Z_{\mu}\left(  x\right)  ,\label{VSL1.12}\\
S_{LV\left(  1\right)  }  &  =ig_{2}^{\prime}\sin\theta\int d^{4}x\left[
j_{1}^{\eta}(x)C^{\mu}\partial_{\eta}Z_{\mu}\left(  x\right)  \right]  ,\\
S_{LV\left(  2\right)  }  &  =-ig_{2}^{\prime}\sin\theta\int d^{4}x\left[
j_{1}^{\mu}(x)C^{\lambda}\partial_{\lambda}Z_{\mu}\left(  x\right)  \right]  .
\end{align}
In order to evaluate these elements, we propose plane wave expansions,
$Z_{\mu}^{0}\left(  x\right)  =N_{k}\varepsilon_{\mu}\left(  k,\lambda\right)
\exp\left(  -ik\cdot x\right)  ,$ $\psi_{l}\left(  x\right)  =N_{q}%
u_{l}\left(  q,s\right)  \exp\left(  -iq\cdot x\right)  ,$ $\psi_{\bar{l}%
}\left(  x\right)  =N_{q^{\prime}}v\left(  q^{\prime},s^{\prime}\right)
\exp\left(  iq^{\prime}\cdot x\right)  ,$ where $k,q,q^{\prime}$ stand for the
4-momentum of the $Z^{0}$ boson and the emerging leptons, respectively, and
$N_{q}=\left(  2Vq_{0}\right)  ^{-1/2}.$ With these expressions, we obtain%
\begin{align}
S_{0}  &  =i\frac{g}{4\cos\theta}\left(  2\pi\right)  ^{4}\frac{\delta
^{4}\left(  q+q^{\prime}-k\right)  }{\left[  8V^{3}q_{0}q_{0}^{\prime}%
k_{0}\right]  ^{1/2}}M_{0},\label{S0}\\
S_{LV\left(  a\right)  }  &  =\frac{ig_{2}^{\prime}\sin\theta}{2}\left(
2\pi\right)  ^{4}\frac{\delta^{4}\left(  q+q^{\prime}-k\right)  }{\left[
8V^{3}q_{0}q_{0}^{\prime}k_{0}\right]  ^{1/2}}M_{LV\left(  a\right)  },
\end{align}
with $a=1,2$ representing the two LV contributions, which involves
\begin{align}
&  \left.  M_{0}=\varepsilon_{\mu}\left(  k,\lambda\right)  \bar{u}_{l}\left(
q,s\right)  \gamma^{\mu}\left(  g_{V}^{\prime}-\gamma_{5}\right)  v\left(
q^{\prime},s^{\prime}\right)  ,\right. \\
&  \left.  M_{LV\left(  1\right)  }=C^{\mu}k_{\eta}\varepsilon_{\mu}\left(
k,\lambda\right)  j_{qq^{\prime}}^{\eta},\right. \label{MLV1a}\\
&  \left.  M_{LV\left(  2\right)  }=-C^{\lambda}k_{\lambda}\varepsilon_{\mu
}\left(  k,\lambda\right)  j_{qq^{\prime}}^{\mu},\right.  \label{MLV1b}%
\end{align}
and $j_{qq^{\prime}}^{\mu}=\bar{u}_{l}\left(  q,s\right)  \gamma^{\mu}\left(
3+\gamma_{5}\right)  v\left(  q^{\prime},s^{\prime}\right)  .$ The decay rate
for the process (\ref{process1}) is given as usually evaluated, that is,
\begin{equation}
\Gamma_{ll}=\frac{1}{T}V\int\frac{d^{3}q}{\left(  2\pi\right)  ^{3}}V\int%
\frac{d^{3}q^{\prime}}{\left(  2\pi\right)  ^{3}}\frac{1}{3}\sum_{\lambda}%
\sum_{s,s^{\prime}}\left\vert S\right\vert ^{2}, \label{VSL1.18}%
\end{equation}
where $S$ is given in (\ref{VSL1.11}), implying
\begin{equation}
\left\vert S\right\vert ^{2}=S_{0}S_{0}^{\dagger}+S_{0}S_{LV\left(  1\right)
}^{\dagger}+S_{LV\left(  1\right)  }S_{0}^{\dagger}+S_{0}S_{LV\left(
2\right)  }^{\dagger}+S_{LV\left(  2\right)  }S_{0}^{\dagger}, \label{S2}%
\end{equation}
in first order in the LV parameters.\textbf{ }Substituting Eq. (\ref{S2}) in
Eq. (\ref{VSL1.18}), we achieve%
\begin{align}
\Gamma_{ll}  &  =\Gamma_{S_{0}S_{0}^{\dagger}}+\Gamma_{S_{0}S_{LV\left(
1\right)  }^{\dagger}}+\Gamma_{S_{LV\left(  1\right)  }S_{0}^{\dagger}%
}\nonumber\label{VSL1.19.2}\\
&  +\Gamma_{S_{0}S_{LV\left(  2\right)  }^{\dagger}}+\Gamma_{S_{LV\left(
2\right)  }S_{0}^{\dagger}}.
\end{align}
The first term, $\Gamma_{S_{0}S_{0}^{\dagger}},$ is the decay rate for the
Lorentz invariant usual process $Z_{0}\rightarrow\bar{l}+l$. In this
evaluation, $\Gamma_{S_{0}S_{LV\left(  1\right)  }^{\dagger}}=0,$
$\Gamma_{S_{LV\left(  1\right)  }S_{0}^{\dagger}}=0,$ as a consequence of the
current conservation, due to the presence of the momentum $k_{\alpha}$ in Eqs.
(\ref{MLV1a}, \ref{MLV1b}). The LV contribution is associated with
$\Gamma_{S_{0}S_{LV\left(  2\right)  }^{\dagger}}$and $\Gamma_{S_{LV\left(
2\right)  }S_{0}^{\dagger}},$ so that the total decay rate, $\Gamma
=\Gamma_{S_{0}S_{0}^{\dagger}}+\Gamma_{S_{0}S_{LV\left(  2\right)  }^{\dagger
}}+\Gamma_{S_{LV\left(  2\right)  }S_{0}^{\dagger}},$ is%

\begin{align}
\Gamma_{ll}  &  =\frac{g^{2}}{1536\pi\cos^{2}\theta}8M_{Z}\left[  \left(
g_{V}^{\prime2}+1\right)  -6g_{V}^{\prime2}\frac{m_{l}^{2}}{M_{Z}^{2}}\right.
\nonumber\\
&  -\frac{gg_{2}^{\prime}\tan\theta}{768\pi}8M_{Z}\left(  C\cdot k\right)
\left[  \left(  3g_{V}^{\prime}-2\right)  -27g_{V}^{\prime}\frac{m_{l}^{2}%
}{M_{Z}^{2}}\right] \nonumber\\
&  \times\Theta\left(  M_{Z}-2m_{l}\right)  . \label{decayZ}%
\end{align}
We now use $k^{2}=M_{Z}^{2}$ and $C\cdot k=C_{0}M_{Z}$. As the $Z_{0}$ mass
($M_{Z}=9.1\times10^{10}eV)$ is much larger than lepton masses, we can neglect
the mass ratios for the electron, muon and tau ($m_{e}^{2}/M_{Z}^{2}%
\simeq2\times10^{-11},$ $m_{\mu}^{2}/M_{Z}^{2}\simeq10^{-6},m_{\tau}^{2}%
/M_{Z}^{2}\simeq4\times10^{-4}),$ which are smaller than the experimental
uncertainty in decay rate measurements. Thus, the result is written as%
\begin{align}
\Gamma_{ll}  &  =\frac{g^{2}\left(  g_{V}^{\prime2}+1\right)  M_{Z}}%
{192\pi\cos^{2}\theta}\left[  1-8\times\left\vert g_{2}^{\prime}%
C_{0}\right\vert M_{Z}\right] \nonumber\\
&  \times\Theta\left(  M_{Z}-2m_{l}\right)  , \label{decayZ2}%
\end{align}
with the LV contribution appearing as a direct correction to the usual decay
rate. We have used $g=e/\sin\theta,$ $g_{V}^{\prime}=1-4\sin^{2}\theta,$
$\sin^{2}\theta=0.23$. In accordance with Ref. \cite{dataZ0}, the $Z_{0}$
decay rate (considering lepton universality) is $\Gamma_{ll}=(83.985\pm
0.086)MeV,$ or $\Gamma_{ll}=83.985(1\pm0.001)MeV,$ so that the experimental
uncertainty is of 1 part in $10^{3}$\textbf{. }We thus impose $8\left\vert
g_{2}^{\prime}C_{0}\right\vert M_{Z}<1.0\times10^{-3},$ which leads to the
following upper bound:%

\begin{equation}
\left\vert g_{2}^{\prime}C_{0}\right\vert <1.3\times10^{-15}\text{ }\left(
eV\right)  ^{-1}. \label{bound1}%
\end{equation}

\section{A nonminimal \textbf{coupling in the SU}$\left(  2\right)  _{L}$
sector of the GSW model}

Analogously to the previous case, a gauge invariant nonminimal interaction in
the $SU\left(  2\right)  _{L}$ sector of the GSW model can be proposed as
\begin{equation}
D_{\mu}=\partial_{\mu}-ig\mathbf{T\cdot W}_{\mu}-i\frac{g^{\prime}}{2}YB_{\mu
}+ig_{3}^{\prime}\left(  \mathbf{T\cdot W}_{\mu\nu}\right)  V^{\nu}.
\label{Dmn1}%
\end{equation}
where $V^{\nu}$ is a fixed 4-vector that establishes a preferred direction in
spacetime and violates Lorentz symmetry. The interaction term, $\bar{L}%
_{l}\gamma^{\mu}i\left(  ig_{3}^{\prime}\mathbf{T\cdot W}_{\mu\nu}V^{\nu
}\right)  L_{l}$ embraces the following interactions at tree-level,
$\mathcal{L}_{LV\left(  2\right)  }=\mathcal{J}_{+\left(  LV\right)
}^{\left(  l\right)  \nu}W_{\nu}^{\left(  +\right)  }+\mathcal{J}_{-\left(
LV\right)  }^{\left(  l\right)  \nu}W_{\nu}^{\left(  -\right)  }%
+\mathcal{J}_{0\left(  LV\right)  }^{\left(  l\right)  \nu}Z_{\nu},$ involving
the vector bosons, in which the related currents read%
\begin{align}
\mathcal{J}_{+\left(  LV\right)  }^{\left(  l\right)  \nu}  &  =-\frac
{g_{3}^{\prime}}{2\sqrt{2}}\bar{\psi}_{\nu_{l}}\gamma^{\mu}\left(
1-\gamma_{5}\right)  \psi_{l}V^{\nu}\partial_{\mu}\nonumber\\
&  +\frac{g_{3}^{\prime}}{2\sqrt{2}}\bar{\psi}_{\nu_{l}}\gamma^{\nu}\left(
1-\gamma_{5}\right)  \psi_{l}V^{\mu}\partial_{\mu},\label{JW1}\\
\mathcal{J}_{-\left(  LV\right)  }^{\left(  l\right)  \nu}  &  =-\frac
{g_{3}^{\prime}}{2\sqrt{2}}\bar{\psi}_{l}\gamma^{\mu}\left(  1-\gamma
_{5}\right)  \psi_{\nu_{l}}V^{\nu}\partial_{\mu}\nonumber\\
&  +\frac{g_{3}^{\prime}}{2\sqrt{2}}\bar{\psi}_{l}\gamma^{\nu}\left(
1-\gamma_{5}\right)  \psi_{\nu_{l}}V^{\mu}\partial_{\mu}, \label{J-W}%
\end{align}%
\begin{gather}
\mathcal{J}_{0\left(  LV\right)  }^{\left(  l\right)  \nu}=-\frac
{g_{3}^{\prime}\cos\theta}{4}\left\{  \bar{\psi}_{\nu_{l}}\gamma^{\mu}\left(
1-\gamma_{5}\right)  \psi_{\nu_{l}}V^{\nu}\partial_{\mu}\right. \nonumber\\
-\bar{\psi}_{l}\gamma^{\mu}\left(  1-\gamma_{5}\right)  \psi_{l}V^{\nu
}\partial_{\mu}-\bar{\psi}_{\nu_{l}}\gamma^{\nu}\left(  1-\gamma_{5}\right)
\psi_{\nu_{l}}V^{\mu}\partial_{\mu}\nonumber\\
\left.  +\bar{\psi}_{l}\gamma^{\nu}\left(  1-\gamma_{5}\right)  \psi_{l}%
V^{\mu}\partial_{\mu}\right\}  .
\end{gather}

The current, $J_{-\left(  LV\right)  }^{\left(  l\right)  \mu}$, given by Eq.
(\ref{J-W}), affects the processes mediated by the $W^{-}$ particle, including
the decay $W^{-}\rightarrow l+\bar{\nu}_{l}$. The total electroweak current
that contributes to this process is%
\begin{gather}
\left(  J_{-}^{\left(  l\right)  \mu}+\mathcal{J}_{-\left(  LV\right)
}^{\left(  l\right)  \mu}\right)  W_{\mu}^{\left(  -\right)  }=\frac{1}%
{2\sqrt{2}}\left[  gj_{2}^{\mu}W_{\mu}^{\left(  -\right)  }\left(  x\right)
\right. \nonumber\\
\left.  -g_{3}^{\prime}[j_{2}^{\eta}V^{\mu}\partial_{\eta}W_{\mu}^{\left(
-\right)  }\left(  x\right)  ]+g_{3}^{\prime}[j_{2}^{\mu}V^{\lambda}%
\partial_{\lambda}W_{\mu}^{\left(  -\right)  }\left(  x\right)  ]\right]  ,
\end{gather}
where $j_{2}^{\mu}(x)=\bar{\psi}_{l}\left(  x\right)  \gamma^{\mu}\left(
1-\gamma_{5}\right)  \psi_{\nu_{l}}$, the first term is the usual Lorentz
invariant contribution. The scattering matrix for the process $\left(
W^{-}\rightarrow l+\bar{\nu}_{l}\right)  ,$ at leading order, can be written
as
\begin{equation}
\mathcal{S}=-i\int d^{4}x\left(  J_{-}^{\left(  l\right)  \mu}+\mathcal{J}%
_{-\left(  LV\right)  }^{\left(  l\right)  \mu}\right)  W_{\mu}^{\left(
-\right)  },
\end{equation}
that implies $\mathcal{S}=\mathcal{S}_{0}+\mathcal{S}_{LV\left(  1\right)
}+\mathcal{S}_{LV\left(  2\right)  },$ with
\begin{align}
\mathcal{S}_{0}  &  =-i\frac{g}{2\sqrt{2}}\int d^{4}x\text{ }\left[
j_{2}^{\mu}(x)W_{\mu}^{\left(  -\right)  }\left(  x\right)  \right]  ,\\
\mathcal{S}_{LV\left(  1\right)  }  &  =i\frac{g_{3}^{\prime}}{2\sqrt{2}}\int
d^{4}x\text{ }\left[  j_{2}^{\eta}\left(  x\right)  V^{\mu}\partial_{\eta
}W_{\mu}^{\left(  -\right)  }\left(  x\right)  \right]  ,\\
\mathcal{S}_{LV\left(  2\right)  }  &  =-i\frac{g_{3}^{\prime}}{2\sqrt{2}}\int
d^{4}x\text{ }\left[  j_{2}^{\mu}\left(  x\right)  V^{\lambda}\partial
_{\lambda}W_{\mu}^{\left(  -\right)  }\left(  x\right)  \right]  .
\end{align}
Following the same steps of previous calculation, we obtain the decay rate for
the usual Lorentz invariant process $\left(  W^{-}\rightarrow l+\bar{\nu}%
_{l}\right)  $:%
\begin{equation}
\Gamma_{\mathcal{S}_{0}\mathcal{S}_{0}^{\dagger}}=\frac{g^{2}}{48\pi}%
M_{W}\left(  1-\frac{m_{l}^{2}}{M_{W}^{2}}\right)  ^{2}\left(  1+\frac
{m_{l}^{2}}{2M_{W}^{2}}\right)  \Theta\left(  M_{W}-m_{l}\right)  ,
\label{W18}%
\end{equation}
where $M_{W},m_{l}$ label the $W^{-}$ boson and lepton masses. As it occurs in
the previous case, the quantities $\Gamma_{\mathcal{S}_{0}\mathcal{S}%
_{LV\left(  1\right)  }^{\dagger}},$ $\Gamma_{\mathcal{S}_{LV\left(  1\right)
}\mathcal{S}_{0}^{\dagger}}$ also vanish, $\Gamma_{\mathcal{S}_{0}%
\mathcal{S}_{LV\left(  1\right)  }^{\dagger}}=0,$ $\Gamma_{\mathcal{S}%
_{LV\left(  1\right)  }\mathcal{S}_{0}^{\dagger}}=0.$ The terms,
$\Gamma_{\mathcal{S}_{0}\mathcal{S}_{LV\left(  2\right)  }^{\dagger}}%
,\Gamma_{\mathcal{S}_{LV\left(  2\right)  }\mathcal{S}_{0}^{\dagger}}$ are
computed, leading to the following decay rate:%

\begin{equation}
\Gamma=\left[  \frac{g^{2}}{48\pi}M_{W}+\left(  g_{3}^{\prime}V_{0}\right)
\frac{10gM_{W}^{2}}{384\pi}\right]  \Theta(M_{W}-m_{l}),
\end{equation}
where $V\cdot k=V_{0}M_{W}$ for the rest frame of the $W^{-}$ mediator, and we
have neglected the contributions in $m_{l}^{2}/M_{W}^{2},m_{l}^{4}/M_{W}^{4}$.
This result can be also expressed as%
\begin{equation}
\Gamma=\frac{g^{2}}{48\pi}M_{W}\left[  1+\left(  g_{3}^{\prime}V_{0}\right)
\frac{5M_{W}}{4g}\right]  \Theta(M_{W}-m_{l}). \label{decayW2}%
\end{equation}
\textbf{ }Considering that the experimental uncertainty in the measures of
this decay is at the level of\textbf{ }$\sim4.0\times10^{-2},$ and using
$g=e/\sin\theta,$\textbf{ }$\sin^{2}\theta=0.23,$ we impose $7\left(
g_{3}^{\prime}V_{0}\right)  M_{W}<4.0\times10^{-2},$\textbf{ }which implies%
\begin{equation}
\left\vert g_{3}^{\prime}V_{0}\right\vert <7\times10^{-14}\text{ }\left(
eV\right)  ^{-1}. \label{bound2}%
\end{equation}
As the current (\ref{JW1}), involving the mediator $W^{+},$ is analogue to the
current (\ref{J-W}), we conclude that these latter developments equally hold
to the decay $W^{+}\rightarrow\bar{l}+\nu_{l},$ which becomes constrained by a
bound similar to Eq. (\ref{bound2}).

\section{Conclusion and final remarks}

In this investigation, we have proposed two CPT-odd nonminimal couplings in
the Electroweak sector, one in the $U\left(  1\right)  _{Y}$ sector, another
one in the $SU\left(  2\right)  _{L}$ sector of the GSW theory. The NM
interaction terms we found at the level of the lepton Lagrangian. The implied
corrections to the decay rates of the processes, $Z_{0}\rightarrow\bar{l}+l$
and $W^{-}\rightarrow l+\bar{\nu}_{l}$ ($W^{+}\rightarrow\bar{l}+\nu_{l}),$
were explicitly computed. Regarding the experimental imprecision in the
measurements, we have impose upper limits on the magnitude of the LV
nonminimal coupling at the level of $10^{-15}$ (eV)$^{-1}$ and $10^{-14}$
(eV)$^{-1}.$

In theories in which the Lorentz symmetry is broken by fixed backgrounds in
spacetime, the tensor components are usually considered fixed in the Sun's
frame, in such a way that they undergo sidereal variations in the Earth frame
\cite{Jonas1,Sideral}. It is necessary, therefore, to translate the bounds
from the Earth-located Lab's RF at the colatitude $\chi$, rotating around the
Earth's axis with angular velocity $\Omega$, to the Sun%
\'{}%
s frame. For experiments up to a few weeks long, the transformation law for a
rank-1 tensor, $A_{\mu}$, is merely a spatial rotation,\textbf{ }$A_{\mu}%
^{T}=\mathcal{R}_{\mu\alpha}A_{\alpha},$ where the label $T$ indicates the
quantity measured in the Sun's frame, and $\mathcal{R}_{0i}=\mathcal{R}%
_{i0}=0$ and $\mathcal{R}_{00}=1.$ Thus, four vector time-components are not
modified,$\ \ A_{0}^{T}=A_{0}$, so that the upper bounds (\ref{bound1}),
(\ref{bound2}) could be equally written in the Sun's frame. However, the
situation is not so simple, as pointed out in Ref. \cite{Pion} (for pion
decays), once the decay rates (\ref{decayZ2}), (\ref{decayW2}) were carried
out in the rest frame of the decaying vector bosons, not in the Lab (Earth)
frame, where the measurements are performed. In order to take into account
this point, one option is to translate the upper bounds (\ref{bound1}),
associated with an evaluation at the vector boson rest frame, directly to the
Sun%
\'{}%
s frame, with the boost
\begin{equation}
C^{0}=\gamma_{z}(C_{T}^{0}+\alpha^{i}C_{T}^{i}),
\end{equation}
where $\gamma_{z}=\gamma(\mathrm{v}_{z})$ is the Lorentz factor,
$\mathrm{v}_{z}$ is the boson velocity in the Sun%
\'{}%
s frame, $\alpha^{i}=\mathrm{v}_{z}^{i}$ $/c$. The data about width decays
were attained in the LEP accelerator \cite{dataZ0}, constructed to work with
centre-of-mass energy around 91 GeV, reaching 161 GeV in 2000. As the $Z_{0}$
mass is close to the centre-of-mass energy, it happens that the Lorentz factor
is nearly 1 ($\gamma_{z}\gtrsim1),$ and not larger than 2, which also implies
a not (meaningful) relativistic velocity ($\mathrm{v}_{z}^{i})$. In this case,
the upper bounds (\ref{bound1}), (\ref{bound2}) can be read in the Sun%
\'{}%
s frame as%

\begin{align}
\left\vert g_{2}^{\prime}(C_{T}^{0}+\alpha^{i}C_{T}^{i})\right\vert  &
\lesssim1\times10^{-15}\text{ }\left(  eV\right)  ^{-1},\label{boundS1}\\
\left\vert g_{3}^{\prime}(V_{T}^{0}+\alpha^{i}V_{T}^{i})\right\vert  &
\lesssim1\times10^{-14}\text{ }\left(  eV\right)  ^{-1}. \label{boundS2}%
\end{align}
If the case the centre-of-mass energy is really close to the boson mass
($\gamma_{z}\simeq1),$ these bounds simplify to $\left\vert g_{2}^{\prime
}C_{T}^{0}\right\vert <1\times10^{-15}$ $\left(  eV\right)  ^{-1}$ and
$\left\vert g_{3}^{\prime}V_{T}^{0}\right\vert <1\times10^{-14}$ $\left(
eV\right)  ^{-1}.$ Another possibility is to write the results (\ref{decayZ2}%
), (\ref{decayW2}) in the Lab frame, in which $C\cdot k=M_{z}(C_{0}-\gamma
_{z}\alpha^{i}C^{i})$, $V\cdot k=M_{z}(V_{0}-\gamma_{z}\alpha^{i}V^{i}),$
procedure that recovers the bounds (\ref{boundS1}), (\ref{boundS2}) for
$\gamma_{z}\simeq1.$

Finally, other impacts of these NMC can be investigated, with attention to the
possible evaluation of differential decay rates of polarized processes, which
could, in principle, yield improved upper bounds by two orders of magnitude.

\begin{acknowledgments}
The authors are grateful to J.A. Helayel-Neto and M. Schreck for useful
comments. They also acknowledge CNPq, CAPES and FAPEMA (Brazilian research
agencies) for invaluable financial support.
\end{acknowledgments}

\end{document}